\documentclass[conference]{IEEEtran}
\IEEEoverridecommandlockouts
\usepackage{cite}
\usepackage{amsmath,amssymb,amsfonts}
\usepackage{hyperref}
\usepackage{algorithmic}
\usepackage{graphicx}
\usepackage{textcomp}
\usepackage{xcolor}
\usepackage{svg}
\usepackage{siunitx} 

\def\BibTeX{{\rm B\kern-.05em{\sc i\kern-.025em b}\kern-.08em
    T\kern-.1667em\lower.7ex\hbox{E}\kern-.125emX}}

\usepackage{eso-pic}
\usepackage{url}
\AddToShipoutPictureBG*{
  \AtPageUpperLeft{%
    \put(0,-40){\raisebox{15pt}{\makebox[\paperwidth]{\begin{minipage}{21cm}\centering
      \textcolor{gray}{This article has been accepted for publication in the proceedings of the 2022 IEEE Sensors. \\Content may change prior to final publication. DOI: 10.1109/SENSORS52175.2022.9967240} 
    \end{minipage}}}}%
  }
  \AtPageLowerLeft{%
    \raisebox{25pt}{\makebox[\paperwidth]{\begin{minipage}{21cm}\centering
      \textcolor{gray}{ \copyright 2022 IEEE.  Personal use of this material is permitted.  Permission from IEEE must be obtained for all other uses, in any current or future media, including reprinting/republishing this material for advertising or promotional purposes, creating new collective works, for resale or redistribution to servers or lists, or reuse of any copyrighted component of this work in other works.
      }
    \end{minipage}}}%
  }
}
    
\begin{document}

\title{Machine Learning In-Sensors: Computation-enabled Intelligent Sensors For Next Generation of IoT  \\
}

\author{
    \IEEEauthorblockN{Andrea Ronco, Lukas Schulthess, David Zehnder, Michele Magno}
    \IEEEauthorblockA{\textit{Dep. Information Technology and Electrical Engineering} \\
    \textit{ETH Zürich}\\
    andrea.ronco@pbl.ee.ethz.ch}

}
\vspace{-1cm}

\newcommand{\timesteps}{$N$ }
\newcommand{\subsets}{$L$ }
\newcommand{\stredname}{ISM330AILP }
\newcommand{\st}{STMicroelectronics }

\maketitle

\begin{abstract}
Smart sensors are an emerging technology that allows combining the data acquisition with the elaboration directly on the Edge device, very close to the sensors.
To push this concept to the extreme, technology companies are proposing a new generation of sensors allowing to move the intelligence from the edge host device, typically a microcontroller, directly to the ultra-low-power sensor itself, in order to further reduce the miniaturization, cost and energy efficiency.
This paper evaluates the capabilities of a novel and promising solution from STMicroelectronics. The presence of a floating point unit and an accelerator for binary neural networks provide capabilities for in-sensor feature extraction and machine learning.
We propose a comparison of full-precision and binary neural networks for activity recognition with accelerometer data generated by the sensor itself.
Experimental results have demonstrated that the sensor can achieve an inference performance of 10.7 cycles/MAC, comparable to a Cortex-M4-based microcontroller, with full-precision networks, and up to 1.5 cycles/MAC with large binary models for low latency inference, with an average energy consumption of only \SI{90}{\micro\J}/inference with the core running at 5 MHz. 
\end{abstract}


\section{Introduction}
\label{sec:intro}

Artificial Intelligence and Machine Learning paradigms are increasing their presence in the area of the Internet of Things (IoT) and Smart Sensors to face the challenging task of automatically extracting information from the data generated by "traditional" sensors spread in the physical world \cite{FANN-on-MCU}.
An emerging trend is to bring machine learning on low-power IoT devices, in contrast with the Cloud Computing paradigm, where the data processing is performed in the cloud.
This trend is motivated by the huge amount of data produced by a growing number of IoT devices.
Transferring all the data to the cloud would have several drawbacks including high energy requirements for data transmission, and, more generally, reliability, latency, and privacy concerns \cite{EdgeComputing}\cite{edge_computing_privacy}. 



Major challenges arise when processing data at the Edge, mostly due to the highly constrained resources available on smart IoT devices\cite{Magno2009,Jelicic2010}.
Thus, bringing intelligence to the edge creates fascinating challenges for industrial and academic researchers \cite{EdgeComputing2}.
In fact, most of today's low-power IoT smart sensors are composed of a low-power microcontroller (MCU), a set of sensors, and an optional but common wireless interface to transmit the findings of the device to a remote gateway or backend \cite{ContiWear}. 

Many academic and sensor technology companies, including Bosch, TDK, \st and many others, are proposing a new generation of smart sensors that embed ultra-low power digital signal processors in the same integrated circuit.
In particular, there is interest in combining the sensing part with a small neural network core to process the sensor data on the fly, increasing the overall efficiency and reducing the system latency.

This novel approach is based on the intuition that waking up external processors periodically might have an impact on the power consumption, which is avoidable by moving the intelligence from the host processor to the sensor itself.
This paper presents the evaluation of a novel sensor with an Intelligent Sensor Processing Unit (ISPU) designed and produced by \st. The ISPU includes both full precision and binary neural network accelerators as well as the capability of DSP for feature extraction of the raw data. The main contribution of this paper is the design of benchmark neural networks (both full precision and binary) to evaluate the new \stredname sensor from \st under different loads to characterize the performance, including the features extraction on the ISPU, and the usability of this solution.
We compare different neural network architectures using both full-precision and binary implementations, with the aim to evaluate the power efficiency of the core in different conditions.
Experimental evaluation in terms of Multiply-Accumulate (MAC) operations per cycle and power in different conditions has been performed to have a fair comparison with a popular microcontroller for smart sensors based on ARM Cortex-M4F.
\section{STRed Intelligent Sensor Processing Unit}
\label{sec:stred}
\begin{figure}
    \centering
    \includegraphics[width=.85\linewidth]{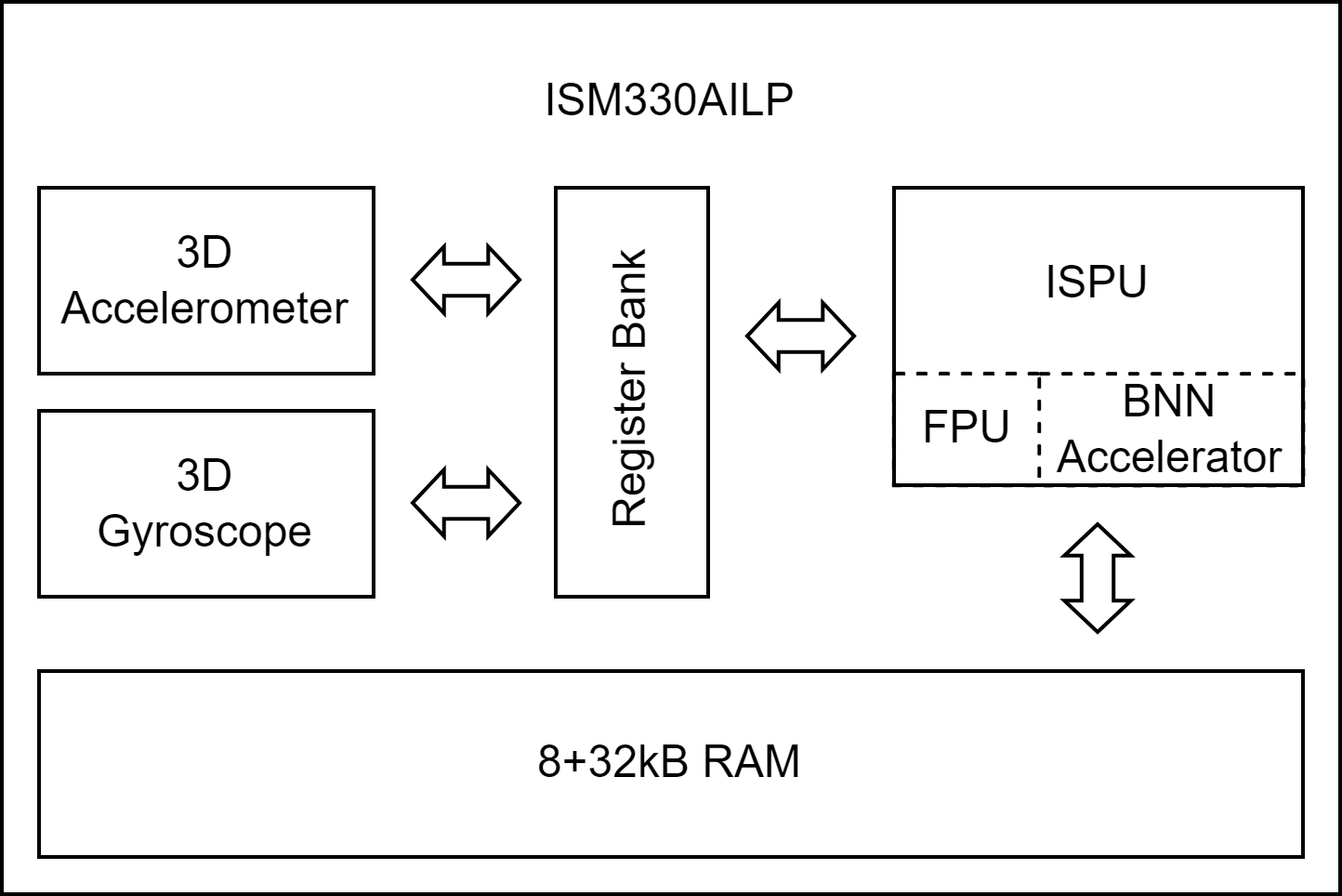}
    \caption{High-level block diagram of the \stredname}
    \vspace{-0.5cm}
    \label{fig:stred_diagram}
\end{figure}

The \stredname inertial module is composed of a low-power, always-on, 6-axis Inertial Measurement Unit (IMU) and the novel Intelligent Sensor Processing Unit (ISPU), as represented in Figure \ref{fig:stred_diagram}.
The main novelty of this sensor is the presence of a fully programmable \SI{32}{\bit} core inside the sensor itself.
The processing core (ISPU) is optimized for power consumption and machine learning, and features \SI{40}{\kibi\byte} of RAM memory (for both program and data), a Floating Point Unit (FPU) for full-precision operations, and an additional Binary Neural Network accelerator with support for convolutions, to enable efficient BNN execution in the sensor.
The sensor offers a high-performance mode and a low-power one, with different sampling frequency ranges, while the ISPU can run at \SI{5}{\mega\hertz} or at \SI{10}{\mega\hertz}.

While the processing core allows the implementation of any algorithm, compatibly with the computational resources, the ISPU is designed with a focus on Binary Neural Networks, which are a competitive alternative to full-precision models \cite{bnn_good}, especially when coupled with ad-hoc accelerators \cite{bnn_yoda} \cite{bnn_xor}.

The core has been designed with a wake-up time of only \num{4} cycles, to allow efficient event-based processing with minimal latency. It wakes up when new sensor data are available, but it is also possible to trigger the execution of algorithms from the external host MCU if needed.

The core can be programmed in C with user-defined algorithms, or with the support of the tools provided by \st , which allow to deploy Keras \cite{keras} models and take advantage of the BNN accelerator.

\section{Evaluation Scenario and proposed neural network}
To evaluate the potential of in-sensor intelligence with the ISPU under a realistic real-world setting, we chose an activity recognition task.
Specifically, we opted for an activity classification of different actions on an office chair, as this allows us to use directly the data from the accelerometer hosted in the same die of the ISPU. 
The set of activities we wanted to recognize is the following:
\begin{itemize}
    \item \textit{Idle}: no motion is detected on the chair.
    \item \textit{Stand up}: a person stands up leaving the chair empty.
    \item \textit{Sit down}: a person sits on an empty chair.
    \item \textit{Rotate}: the chair spins around the shaft.
    \item \textit{Move}: the chair is moved around.
\end{itemize}

\begin{figure}[h]
    \centering
    \includegraphics[width=0.9\linewidth]{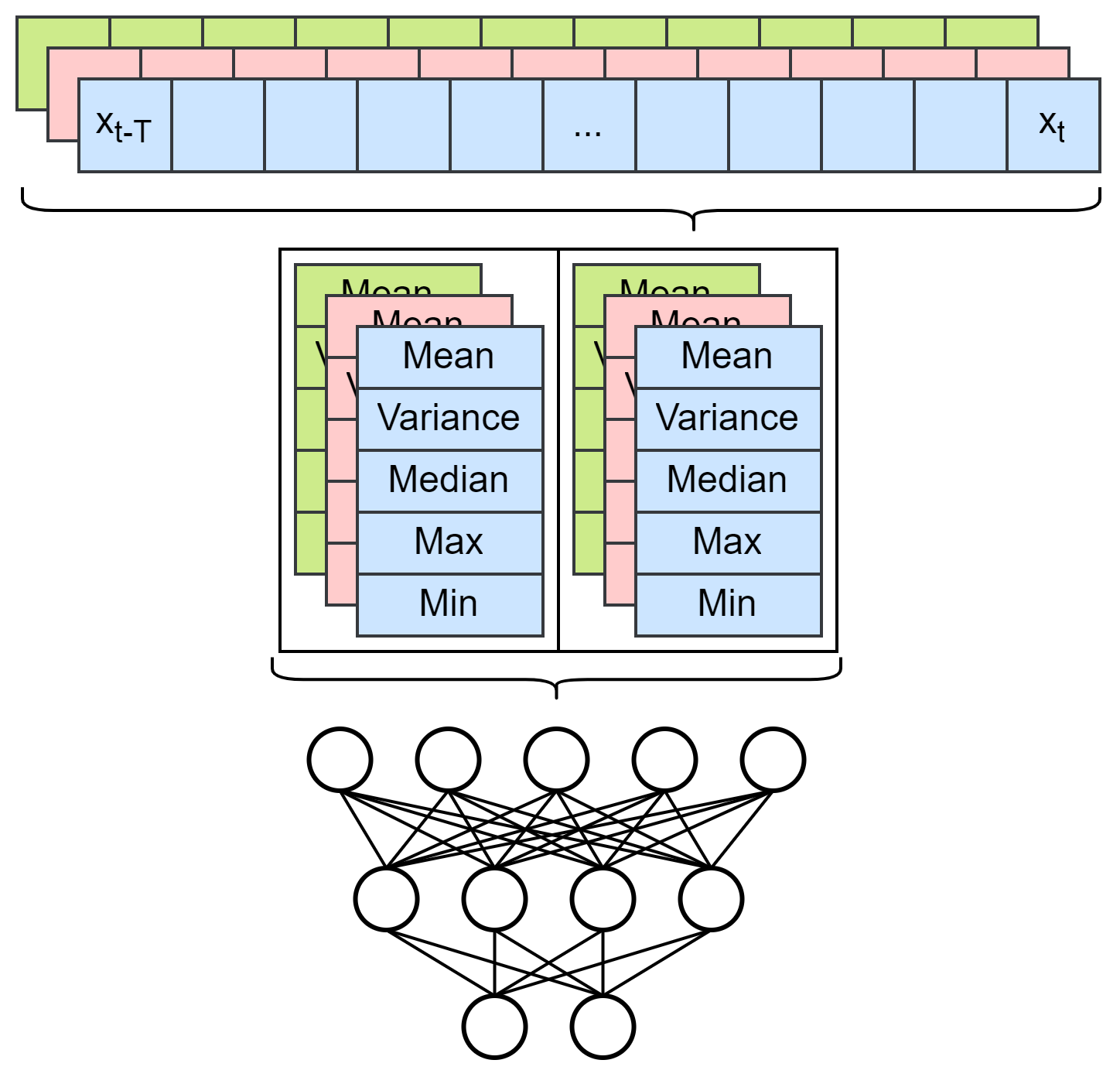}
    \caption{Graphical representation of our classification pipeline. The network architecture is not representative.}
    \vspace{-0.5cm}
    \label{fig:pipeline}
\end{figure}

For our demonstration, we consider a prediction pipeline composed of a \textit{Feature Extractor} followed by a \textit{Neural Network}, as shown in Figure \ref{fig:pipeline}.
This well-known combination \cite{features_nn} allows us to evaluate the performance of the ISPU for both traditional feature extraction/signal processing tasks and machine learning loads.

\subsection{Feature Extraction}
We adopted a feature extraction algorithm based on standard time-domain statistical features, which are suitable for real-time systems \cite{features} and are invariant to the sampling rate of the sensor.
Since the ISPU is triggered at every acquisition, we consider 1 acquisition as our time unit to simplify the explanation below.
Every acquisition yields three \SI{16}{\bit} ADC samples, one per axis. The samples are stored in three dedicated circular buffers of length $N=32$.

With a periodicity of $T=32$ (when the buffers are full), we evaluate for each axis the following set of features: \textit{Mean}, \textit{Median}, \textit{Variance}, \textit{Maximum}, \textit{Minimum}.
The set of features $S_t$ is stored in a shift buffer which contains the last $L=2$ sets, from $S_t$ to $S_{t-(L-1)}$.
Together they compose the input of the neural networks described later.

The configuration we used in our application generates a total of \num{30} features, which compose the input for the neural networks described below.

\subsection{Neural Networks}
We compare different neural network architectures using both full-precision and binary implementations, with the aim to evaluate the power efficiency of the ISPU in different conditions. We generated networks with growing complexity by adding hidden layers with \num{32} or \num{64} units.
The full-precision networks have an input size of 30 and the hidden layers have ReLU activation. 
The input of the binary network is padded with two zeros to comply with hardware limitations of the BNN core, which requires the input size of the layers to be multiple of \num{32}. Binary activation is used in the hidden layers. All the models begin with a Batch Normalization layer and terminate with a fully connected layer with SoftMax activation.
The model variations are described in Table \ref{tab:nn}, along with the network complexity measured in Multiply-Accumulate operations.

\begin{table}[h]
\centering
\begin{tabular}{ l r r r }
\hline
 \textbf{Model} & \textbf{Hid. Layers} & \textbf{Hidden Units} & \textbf{MACs}\\ 
 \hline
 Float & 0 &  & 290\\ 
 Float\textsubscript{1,32} & 1 & 32 & 1324\\
 Float\textsubscript{1,64} & 1 & 64 & 2508\\
 Float\textsubscript{2,32} & 2 & 32 & 2412\\
 Float\textsubscript{2,64} & 2 & 64 & 6732\\
 Float\textsubscript{3,32} & 3 & 32 & 3500\\
 Binary & 0 &  & 304\\ 
 Binary\textsubscript{1,32} & 1 & 32 & 1328\\
 Binary\textsubscript{1,64} & 1 & 64 & 2640\\
 Binary\textsubscript{2,32} & 2 & 32 & 2416\\
 Binary\textsubscript{2,64} & 2 & 64 & 6864\\
 Binary\textsubscript{3,32} & 3 & 32 & 3504\\
 Binary\textsubscript{4,256} & 4 & 256 & 208272\\
\end{tabular}
\caption{Network architectures.}
\label{tab:nn}
\end{table}

To train the networks, we acquired a dataset with three different chair types. For each chair, we asked \num{25} different people to perform the required \num{5} activities. The models were trained for \num{100} epochs.
We evaluated the models with standard train/validation/test split, and achieved for all models an accuracy between \SI{96}{\percent} and \SI{98}{\percent} for the full-precision ones, and between \SI{93}{\percent} to \SI{97}{\percent} for the binary versions, confirming the expectations to experience a drop in accuracy due to the extreme quantization.
The models were deployed on the sensor with the tools provided by STMicroelectronics, while the feature extraction was implemented directly in C.

\section{Experimental Results}
\label{sec:results}

\subsection{Execution Time and Power Consumption}
Since the toolchain does not include an accurate performance profiler, we evaluate the execution time by measuring the average duration of the ISPU phase in the power profile on about \num{100} samples per configuration.

The feature extraction has an average duration of \SI{6.57}{\ms}, and is common to all the models.

\begin{figure}[h]
    \centering
    \includegraphics[width=\linewidth]{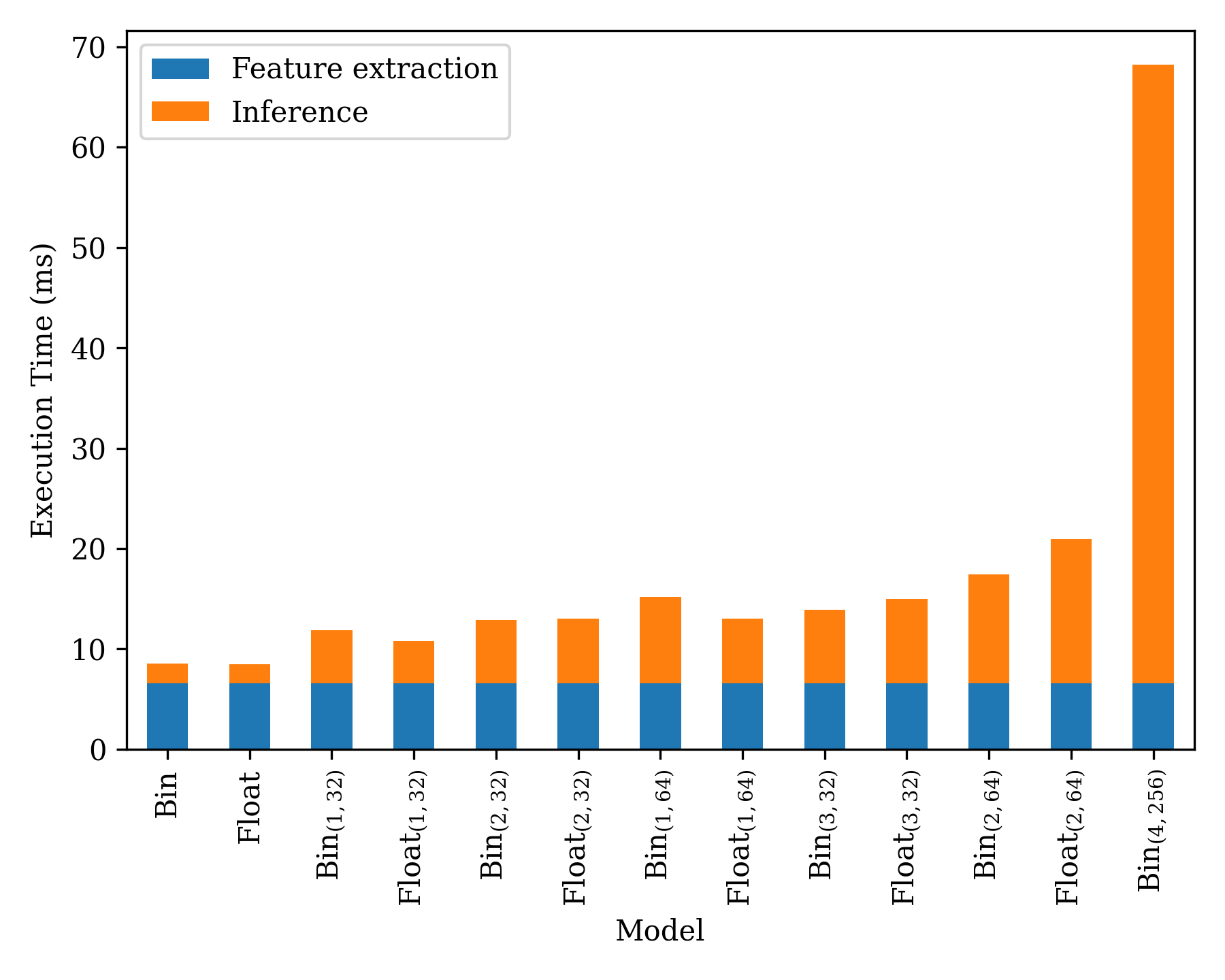}
    \caption{Execution time of different models on the \stredname}
    \vspace{-0.5cm}
    \label{fig:exec_time}
\end{figure}

As visible in Figure \ref{fig:exec_time}, the overhead of the feature extraction is considerable and dominates in the simplest network architectures.
We believe a consistent part of the overhead might be caused by the integer-to-floating-point conversions, which to our knowledge are not optimized.

In Table \ref{tab:flops} we compare the cycles/MAC operation for each network architecture on both the ISPU and an STM32L4R9, an MCU based on the Cortex-M4 core. On the latter, we only evaluate the full-precision models, since the architecture does not have native support for binary networks.\\
The data show that small networks, in general, perform worst than large ones, as the additional overhead of operations such as Batch Normalization/Softmax has a greater impact.
This is even more relevant in the case of binary networks, because of the aforementioned overhead of data conversion needed at the input and the output of the network.
For the same reason the binary models struggle to provide any speedup for small to medium models, granting only about \SI{25}{\percent} speedup for the largest model we could compare ($Float_{(2,64)}$) on the sensor, as visible in Figure \ref{fig:exec_time}.

To test this thesis, we also evaluated a single binary model with \num{4} hidden layers and \num{256} hidden units per layer to maximise the BNN accelerator utilization. While it was not possible to test the full-precision version of this model on the sensor for memory limitations, the binary one achieved 1.48 cycles/MAC operation, a promising speedup of about 7x on the previously measured full-precision performance.

\begin{table}[h]
\centering
\begin{tabular}{ l r r r}
\hline
 \textbf{Model} & \textbf{MACs} & \textbf{Cyc/MAC(M4)} & \textbf{Cyc/MAC(STRed)}\\ 
 \hline
 Float                       & 290    & 18.20 & 32.32\\ 
 Float\textsubscript{1,32}   & 1324   & 13.28 & 15.86\\
 Float\textsubscript{1,64}   & 2508   & 12.03 & 12.76\\
 Float\textsubscript{2,32}   & 2412   & 12.45 & 13.26\\
 Float\textsubscript{2,64}   & 6732   & \textbf{10.91} & \textbf{10.67}\\
 Float\textsubscript{3,32}   & 3500   & 12.13 & 11.98\\
 Binary                      & 304    &       & 32.09\\ 
 Binary\textsubscript{1,32}  & 1328   &       & 19.78\\
 Binary\textsubscript{1,64}  & 2640   &       & 16.32\\
 Binary\textsubscript{2,32}  & 2416   &       & 13.02\\
 Binary\textsubscript{2,64}  & 6864   &       & 7.88\\
 Binary\textsubscript{3,32}  & 3504   &       & 10.46\\
 Binary\textsubscript{4,256} & 208272 &      & \textbf{1.48}\\
\end{tabular}
\caption{Execution time metrics}
\label{tab:flops}
\end{table}


\section{Conclusion}
\label{sec:conclusion}

We proposed an evaluation of on-sensor activity classification with the novel \stredname sensor with integrated ISPU and analyzed the performance of the core under different conditions, proposing neural networks for realistic application scenarios, and exploiting both the sensing and the on-board processor. Our evaluation shows that the ISPU is suitable for running small full-precision networks, and is mainly limited by the available memory for what concerns this task. The cycles/MAC metric also shows that the performance of the core is comparable with the performance of a Cortex-M4 core with intensive full-precision loads with an inference energy of only \SI{90}{\micro\J}. Our experiments also show that it is possible to run large binary models with a speedup of around 7x when using the dedicated BNN accelerator.


\vspace{12pt}

\end{document}